# Laser-Induced Spatially-Selective Tailoring of High-Index Dielectric Metasurfaces

**JONAS BERZINŠ,**[1,2,*] **SIMONAS INDRIŠIŪNAS,**[3] **STEFAN FASOLD,**[1] **MICHAEL STEINERT,**[1] **OLGA ŽUKOVSKAJA,**[4,5] **DANA CIALLA-MAY,**[4,5] **PAULIUS GEČYS,**[3] **STEFAN M. B. BÄUMER,**[2] **THOMAS PERTSCH,**[1,6] **AND FRANK SETZPFANDT**[1]

[1]*Institute of Applied Physics, Abbe Center of Photonics, Friedrich Schiller University Jena, Albert-Einstein-Str. 15, 07745 Jena, Germany*
[2]*TNO Optics Department, Stieltjesweg 1, 2628CK Delft, The Netherlands*
[3]*Department of Laser Technologies, Center for Physical Sciences and Technology, Savanoriu Ave. 231, LT-02300 Vilnius, Lithuania*
[4]*Leibniz Institute of Photonic Technology Jena, Albert-Einstein-Str. 9, 07745 Jena, Germany*
[5]*Institute of Physical Chemistry and Abbe Center of Photonics, Friedrich Schiller University Jena, Helmholtzweg 4, 07745 Jena, Germany*
[6]*Fraunhofer Institute for Applied Optics and Precision Engineering, Albert-Einstein-Str. 7, 07745 Jena, Germany*
**jonas.berzins@uni-jena.de*

**Abstract:** Optically resonant high-index dielectric metasurfaces featuring Mie-type electric and magnetic resonances are usually fabricated by means of planar technologies, which limit the degrees of freedom in tunability and scalability of the fabricated systems. Therefore, we propose a complimentary post-processing technique based on ultrashort ($\leq 10$ ps) laser pulses. The process involves thermal effects: crystallization and reshaping, while the heat is localized by a high-precision positioning of the focused laser beam. Moreover, for the first time, the resonant behavior of dielectric metasurface elements is exploited to engineer a specific absorption profile, which leads to a spatially-selective heating and a customized modification. Such technique has a potential to reduce the complexity in the fabrication of non-uniform metasurface-based optical elements. Two distinct cases, a spatial pixelation of a large-scale metasurface and a height modification of metasurface elements, are explicitly demonstrated.

## 1. Introduction

High-index dielectric metasurfaces, dense planar arrangements of dielectric nanostructures, have been extensively investigated for the unique ability to manipulate properties of light [1,2]. They have been recently used in many demonstrations of optical components, such as filters [3,4], polarizers [5,6], and lenses [7,8], as well as nonlinear phenomena [9–11], and surface enhanced spectroscopy [12,13]. Nevertheless, most of the time, the optically resonant metasurfaces remain inferior to their conventional counterparts based not on the performance, but rather on the size-limited and complex fabrication [14]. Hence, for the metasurface-based elements to be applied in mass-production devices it is necessary to create new fabrication techniques or significantly improve the existing ones.

Silicon (Si) metasurfaces are mainly fabricated by lithographic techniques, such as single-step electron-beam lithography (EBL) and subsequent etching or focused-ion-beam (FIB) lithography [15]. These technologies are very well-developed, but are mostly limited in controlling the lateral geometry of the fabricated elements. Even-though for some applications metasurfaces with uniform height across the whole area are sufficient [16], it imposes a general restriction on the freedom of design and a number of applications could be improved using a variation in height, e.g. the nanostructure-based RGB color filter arrays for digital imaging [17]. Similarly, an improved performance is expected in color printing [18–20], and a variation in height could be used as an additional degree of freedom in the realization of asymmetric metasurfaces [21]. All things considered, the size of lithography systems, the complexity of the fabrication process, and the subsequent costs, are limiting the move to the mass production. Although potentially large-scale nanofabrication techniques exist, such as nanoimprint lithography (NIL) [22], they are optimized for a homogeneous fabrication and could benefit from the possibility to add local functionalities.

An intriguing way to obtain non-uniform metasurfaces is using laser-induced thermal effects [23], where a lithographically pre-processed uniform metasurface is modified by a laser irradiation, as illustrated in **Fig. 1**(a). In comparison to a homogeneous furnace heating [24,25], the laser enables a precise positioning of the heat source, while the induced dominant thermal effects are similar. First, the absorption of the irradiation results in a heating of nanostructures, which after sufficient deposited energy leads to a phase change of the material. The phase change in Si is dynamic to some extent due to its large thermo-optic coefficient [26,27], but becomes permanent at higher temperatures [28]. The crystallization is followed by a drop in the refractive index, thus has a direct impact on the optical response. Moreover, a further increase of the laser fluence initiates reshaping of the nanostructures, governed by a surface diffusion of the molten material to minimize its surface energy [24,25]. As the optical resonances are size- and shape-dependent [29,30], this enables a vivid control of the optical response. Recently, the laser-induced reshaping was experimentally demonstrated in plasmonic [31–38], dielectric [39,40], and hybrid nanostructures [41–43].

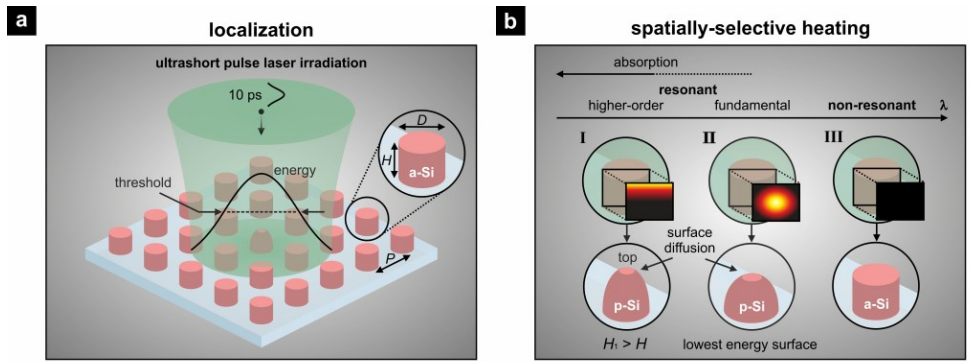

**Fig. 1**. Illustration of laser-induced tailoring of amorphous Si metasurfaces using 10 ps laser pulses. (a) Spatial localization of thermal effects via precise positioning of the focused laser beam. Sufficient laser fluence results in a phase change to polycrystalline phase (p-Si), while at higher temperatures the molten material diffuses to minimize its surface energy. Metasurface elements are defined by diameter $D$, height $H$, and period $P$. (b) Spatially-selective heating engineered by the electromagnetic field distribution in the metasurface elements and the material absorption. Different regimes are shown: (I) high absorption at high order resonances, $\lambda \ll nD$, with hotspot at the top, resulting in taller structures, $H_1 > H$; (II) high absorption at fundamental resonances, $\lambda \approx nD$, with hotspot in the center and reshaping towards lowest energy surface; (III) non-absorptive and non-resonant regime, $\lambda \gg nD$.

In contrast to plasmonics, resonant dielectric metasurfaces possess strong electromagnetic fields inside the volume of its elements [29,30]. Depending on resonant modes and material absorption, this provides a specific field distribution and a successive absorption profile, which can be used for a spatial selectivity of the thermal effects within the metasurface elements, as illustrated in **Fig. 1**(b). However, up to now, this has been barely addressed. Higher order resonances were chosen in melting of Si nanostructures using a CW laser [39] and the resonant behavior was used in a so called resonant laser printing, where the wavelength of 1 ns laser pulse was set at the fundamental resonances of germanium (Ge) metasurfaces [40]. Even-though a strong reshaping was shown in both cases, it was limited to the shape transition from the initial to the lowest energy surface, because of a large thermal diffusion length. Therefore, we propose the use of ultrashort laser pulses to confine the heat and access the full potential of the spatial-selectivity.

In this work we show, how laser-induced thermal effects can be used for flexible and large-scale post-processing of dielectric metasurfaces. In particular, we apply the tailoring on nanostructure-based transmissive spectral filters, implementing a pixelation of a large-scale metasurface. Moreover, we highlight the spatial selectivity of the ultrafast technique to obtain an additional degree of freedom in a spatial control of Si metasurface elements. We analyze the laser-matter interaction by means of geometry analysis, optical and Raman spectroscopies, and complement our results by numerical simulations. We demonstrate the use of ultrashort pulses as a promising complimentary technique for the dielectric metasurfaces beyond the limits of planar fabrication, thus opening new possibilities in the field of nanophotonics.

## 2. Mechanism, Materials and Methods

Laser-induced thermal effects depend on many different parameters: the optical and thermal properties of the target material, the environment it is being modified in, and the parameters of the irradiation source such as spectrum, pulse duration, spatial energy distribution, and energy density (fluence) [44,45]. In addition, nanostructured materials tend to obtain wavelength-dependent optical properties. Si nanostructures, as well as other high-index nanostructures, possess strong fundamental electric dipole (ED) and magnetic dipole (MD) resonances at the wavelengths close to their optical size, $\lambda \approx nD$ [29]. Furthermore, higher-order resonances appear at shorter wavelengths, $\lambda \ll nD$. Each resonant mode is associated with a particular electromagnetic field distribution in the volume of the nanostructures, thus enables a spatial control of energy dissipation. If accounted for, this provides an additional degree of freedom in post-processing of the uniform dielectric metasurfaces.

Our experimental investigation of laser-induced tailoring is based on Si metasurfaces fabricated from a thin film of amorphous Si on top of a glass substrate (see Appendix for dispersion parameters). By means of EBL and reactive ion etching, the film was structured into polarization-insensitive disk-shape nanostructures, introduced in **Fig. 1**. Multiple metasurfaces were fabricated with a diameter $D$ of their elements varying from 55 nm to 165 nm, a period $P$ ranging from 200 nm to 300 nm, while a height $H$ was fixed at 175 nm. The diameter variation within the given range results in metasurfaces with their fundamental ED and MD resonances in a visible (VIS) spectral range, see **Fig. 2**(a). Additionally, a metasurface with the following parameters was investigated: a diameter $D = 560$ nm, a period $P = 794$ nm, and a height $H = 220$ nm, which results in the fundamental resonances in an infrared (IR) region, see **Fig. 2**(d). As the irradiation source we chose a high peak power picosecond laser (Ekspla, Ltd.) with a pulse duration of $\tau_p = 10$ ps, and a pulse repetition rate up to $v = 1024$ kHz. The fundamental wavelength was transformed into a second harmonic, $\lambda = 532$ nm, in order for it to be in the resonant spectral range of the investigated samples as well as the high-absorption of amorphous Si. Before the focus, the diameter of the Gaussian-shape beam was $d_1 = 5$ mm, as measured at $1/e^2$ level of intensity. The laser beam was focused by a $f = 50$ mm lens into a spot-size of $d_0 = 5.6$ µm, calculated by the pulsed beam spot-size measurements [46]. The samples were positioned by a high-precision multi-axis stage (Aerotech, Inc.). The experiments were carried out in ambient conditions.

For the post-processing, first, we overview linear optical properties of the pre-processed samples and their potential influence on the process. In **Fig. 2**, measured transmission spectra and calculated spatial distributions of absorption for two optically distinct Si metasurfaces are presented. In **Fig. 2**(a), we show the transmission spectrum of a metasurface with its resonances in the VIS region ($D = 140$ nm, $P = 250$ nm, $H = 175$ nm). The resonances are denoted by the transmission minimum at $\lambda = 640$ nm, which is associated to the closely spaced ED and MD resonances [4]. In order to gauge, how the energy could be deposited in the metasurface elements, we calculated the spatial distribution of the absorbed power at the wavelength of irradiation, see **Fig. 2**(b). As it is close to the fundamental dipolar resonances, we observe the largest absorption in the middle of the nanostructure. In comparison, irradiation directly at the resonance would yield a qualitatively similar energy distribution – a strong absorption in the middle, see **Fig. 2**(c), as also exploited in the application of resonant laser printing [40]. However, a different behavior is found in the metasurface, which has its fundamental ED and MD resonances in the IR ($D$ = 560 nm, $P$ = 794 nm, $H$ = 220 nm), as shown by the transmission spectrum in **Fig. 2**(d). Here, the fundamental resonances and the excitation wavelength of $\lambda$ = 532 nm are further apart, with the latter coinciding with a high absorption range and higher-order optical resonances. As shown in **Fig. 2**(e), this leads to absorption mostly at the top of the nanostructure, which is substantially different from the previous case as well as the excitation at the fundamental resonances, presented in **Fig. 2**(f).

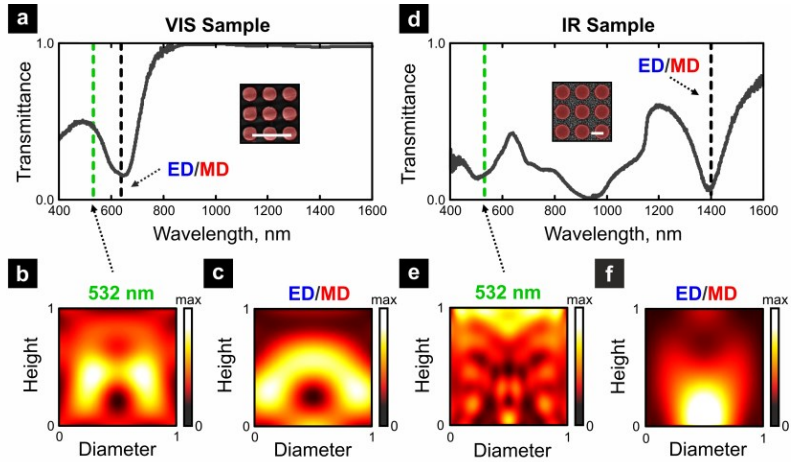

**Fig. 2**. Optical response of Si metasurfaces before a photothermal treatment. (a) Measured transmittance of Si metasurface with ED and MD resonances in the VIS region ($D$ = 140 nm, $P$ = 250 nm, $H$ = 175 nm). (b,c) Calculated absorption profile at the cross section of an element of metasurface from (a), when irradiated at: (b) $\lambda$ = 532 nm, (c) $\lambda$ = 640 nm. (d) Measured transmittance of Si metasurface with ED and MD resonances in the IR ($D$ = 560 nm, $P$ = 794 nm, $H$ = 220 nm). (e,f) Calculated absorption profile at the cross section of an element of metasurface from (d), when irradiated at: (e) $\lambda$ = 532 nm, (f) $\lambda$ = 1400 nm. The insets in (a,d) show colored SEM images of the respective samples, the scale bars are equal to 500 nm.

Our simulations indicate, that the selection of the optical modes enables a spatial control of the onset of the laser-induced thermal effects. The optical simulations were carried out by a finite-difference time-domain method (FDTD, Lumerical, Inc.). The absorbed power at the cross-sections of the metasurface elements shows distinct hot-spots, where the material threshold $F_0$ is reached first, inducing the anticipated modification. The smallest spatial scale is estimated by a thermal diffusion length [47]: $L_{\text{th.}} = (2\alpha\tau_p)^{1/2}$, where $\tau_p$ is a laser pulse duration, and $\alpha$ is a thermal diffusivity, calculated as $\alpha = k/(\rho c_p)$, based on a thermal conductivity $k$, a density $\rho$, and a specific heat $c_p$. As noticed, the length-scale is directly related to the pulse duration: the longer the duration, the larger the thermal diffusion length. Thus, we chose a laser source with a relatively short pulses, $\tau_p = 10$ ps. Using a single pulse of such duration, this allows a spatial resolution as high as ~4 nm, considering that amorphous Si has a density of $\rho = 2280$ kg/m$^3$, a specific heat of $c_p = 880$ J/(kg K), and a relatively low thermal conductivity of $k = 1.8$ W/(m K) [48,49]. In contrast, a longer duration of operation, e.g. using nanosecond pulses or, more so, a CW laser, would induce the heat penetration throughout the whole silicon nanostructure, without the possibility to obtain small features.

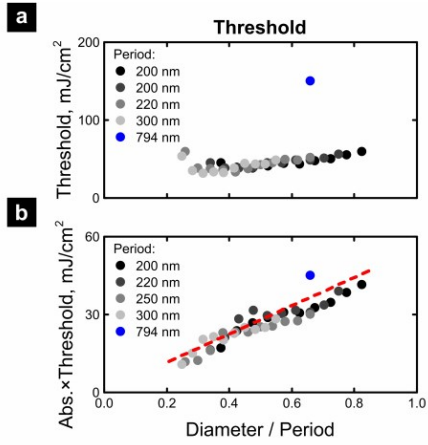

**Fig. 3**. Threshold of Si metasurfaces versus diameter to period ratio: (a) calculated from measured data, according to [46], (b) same values, but normalized to the absorption of the corresponding samples at λ = 532 nm. The samples are color coded by their period: 200 nm, 220 nm, 250 nm, 300 nm, and 794 nm. The threshold grows with the size of the metasurface elements. The dashed line represents a linear fit of the normalized experimental points.

In addition, we perform an experimental study of material threshold $F_0$ for the thermal modifications to take place. It is known that the material damage threshold of amorphous Si thin films is in a range from 100 mJ/cm$^2$ to 220 mJ/cm$^2$ [50–53]. However, there was no in-depth investigation in case of Si metasurfaces and dependence on its geometry. To experimentally determine the threshold $F_0$, we irradiate each investigated metasurface with a set of laser pulses and a stepwise increase of pulse energy. At each step, a diameter $d$ of a modified area is optically characterized, considering that the modification is either intrinsic, if the nanostructure is heated above the crystallization temperature $T_c$, or shape-related, for temperatures close or above the melting temperature $T_m$. We calculate the threshold $F_0$ from the well-defined relation between the diameter $d$ of the modified area and the fluence $F$: $d^2 = d_0^2(\ln F_0 - \ln F)$ [46]. In **Fig. 3**(a), the obtained threshold $F_0$ values are plotted against the diameter to period ratio $D/P$ for all of the investigated samples. There, we indicate that the Si metasurface threshold $F_0$ is similar or smaller to the damage threshold of Si thin film [50–53]. Moreover, the obtained threshold values, normalized to the absorption at λ = 532 nm, follow a linear trend, see **Fig. 3**(b), which is linked to the material surface to air ratio and the high index contrast to the environment. The latter enables a guiding of an incident light beyond the physical size of the metasurface elements, as demonstrated in Si nanowires [54]. The obtained dependence on the diameter to period ratio $D/P$ helps to predict the required fluences for a precise tailoring of the nanostructured samples. Next, we demonstrate two explicit examples, how the laser-induced spatially-selective internal and shape changes of the metasurface elements can be harnessed for controlling their optical response.

## 3. Case Studies

### 3.1 Laser Pixelation of Large-Scale Metasurface

First, we investigate the laser-induced tailoring of amorphous Si metasurfaces with their fundamental ED and MD resonances close to the irradiation wavelength of λ = 532 nm. Due to wavelength-dependent absorption and scattering, such structures can be used as dielectric nanostructure-based transmissive color filters, which were shown as a promising alternative to conventional dye-based filters [3,4]. However, their fabrication is mainly done in a small-scale by EBL. Thus, we propose a technique, called laser pixelation, which, similarly to laser printing [55], allows to use a large-scale uniform fabrication for an initial template and then post-process it for a required pattern, as illustrated in **Fig 5**(a).

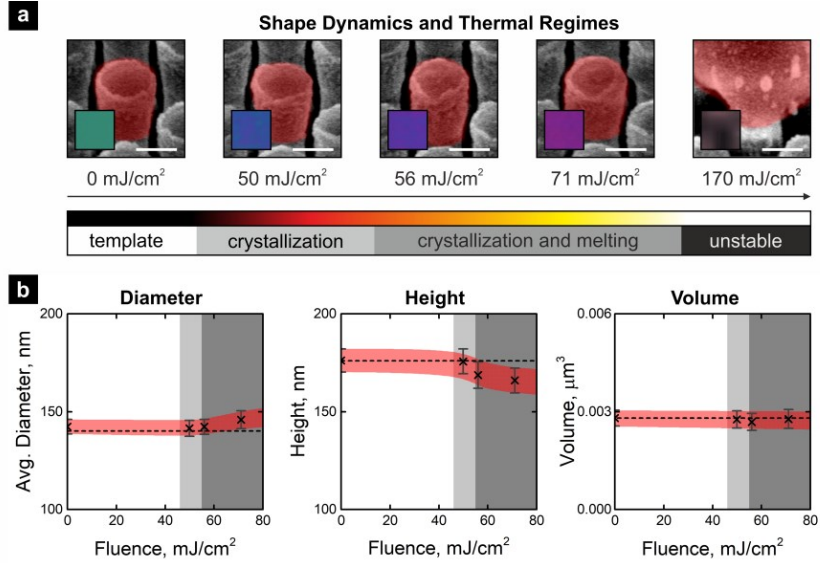

Fig. 4. Shape dynamics due to laser pixelation of Si metasurface elements with ED and MD resonances in the VIS spectral range. (a) SEM images of a unit cell after irradiation with different fluences and respectively identified thermal regimes via shape and Raman measurements. The scale bars denote 100 nm. The insets show the transmitted colors of the metasurfaces. (b) Measured geometrical parameters after irradiation: average diameter (left), height (center), and volume of modified nanostructures (right). The dashed line highlights the values for the unmodified template. Crosses with error bars depict measured values after irradiation, the red shade denotes the interval of uncertainty. The different modified regimes defined in (a), crystallization, crystallization and melting regime, are highlighted in light- and dark grey, respectively.

For the proof of concept, we chose a uniform amorphous Si metasurface ($D$ = 140 nm, $H$ = 175 nm, $P$ = 250 nm) realizing a green (G) color filter, but still possessing absorption inside its elements at the wavelength of irradiation, as shown in **Fig. 2**(b,c). First, a uniform energy distribution over a large area is obtained for a systematic investigation of the laser-induced thermal effects. This was carried out by a partial overlap of several laser shots in a hexagonal configuration, which corresponds to the densest packing of the circular laser spots. The laser fluence at $\lambda$ = 532 nm was increased in several steps until the nanostructures were completely melted and became unstable, i.e. moved from their initial position and disordered the lattice. In **Fig. 4**(a) we show SEM images of experimentally modified metasurface elements after irradiation by fluence $F$ of 50 mJ/cm$^2$, 56 mJ/cm$^2$, 71 mJ/cm$^2$, and 170 mJ/cm$^2$, respectively, together with resulting colors in transmission and approximate thermal regimes, determined by spectroscopic and geometric analysis, visual observations. In general, it is known that the crystallization of amorphous Si starts at $T_c$ = 900 K [28], whereas Si melts at $T_m$ = 1420 K [56]. Furthermore, when the average temperature of the metasurface elements exceeds the melting temperature $T_m$, they melt completely and a chaotic conglomeration is observed, extinguishing the functionality of the metasurface. Even-though a clear threshold for the latter was not distinguished, for this particular sample we predict it around $F = 100$ mJ/cm$^2$. To add, in **Fig. 4**(b) we show measured diameters and heights, as well as calculated volumes of several analyzed metasurface elements at different laser fluences applied. The increase of fluence results in a slight growth of the average diameter $D$ from 142±4 nm to 146±5 nm, while the height $H$ decreases from 176±6 nm to 166±6 nm. As anticipated by the mechanism of dewetting [25], the metasurface elements become rounder at higher energies, but no change is observed in the volume.

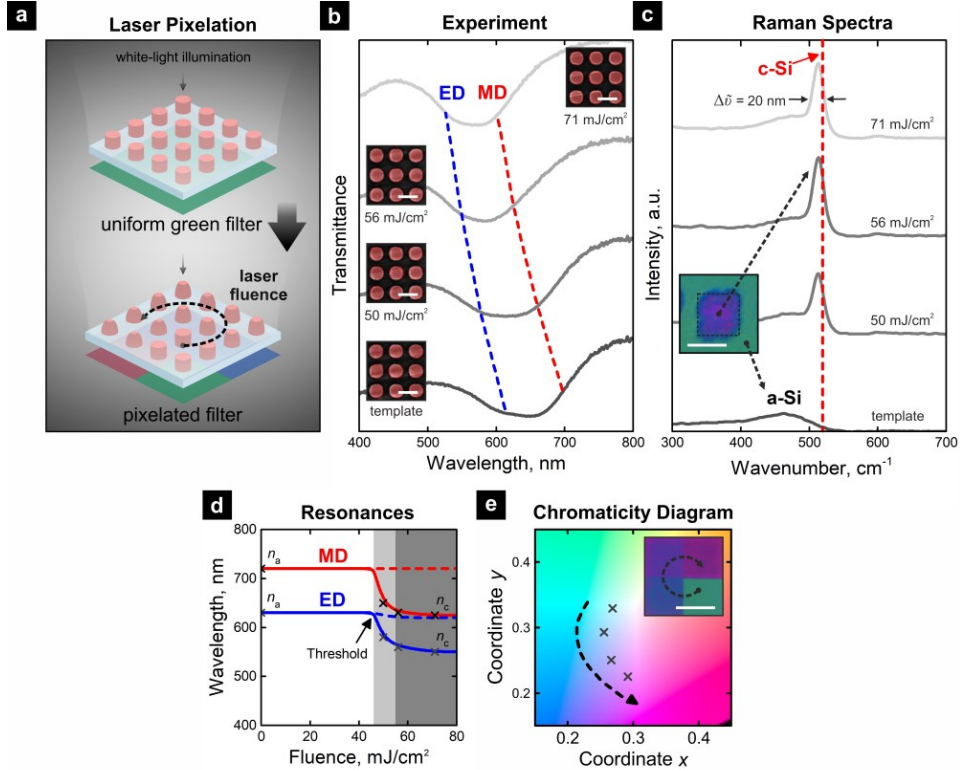

**Fig. 5**. Control of optical resonances based on crystallization. (a) Illustration of laser pixelation, uniform nanostructure-based green (G) filter prior- and post-processing. (b) Transmittance before and after photothermal reshaping of nanostructure-based G color filter. Insets show SEM images, the scale bar is 250 nm. Central wavelengths of ED and MD resonances are highlighted by blue and right dashes lines, respectively (c) Raman spectra measurements of the samples with amorphous Si (a-Si) and crystalline Si (c-Si) fingerprints highlighted. Inset shows a 5 μm pixel with a clear color contrast attributed to the phase change. Scale bar denotes 5 μm. (d) Spectral positions of ED and MD resonances depending on the laser fluence applied. The crosses show points were full analysis was carried out. Dashed lines shows simulations results of amorphous Si with the same geometry. (e) Filter colors realized by modification denoted in the CIE 1931 chromaticity diagram. The inset shows realized multispectral filters from the G filter template, as imaged by transmitted light microscope. The pixel size and scale bar is equal to 50 μm.

Measured transmission spectra of the corresponding metasurfaces are shown in **Fig. 5**(b). In all of the spectra, one can find a dip due the fundamental ED and MD resonances. Although it is not strongly pronounced because of a high absorption of amorphous Si in the VIS spectral range, the resonances and the subsequent dip blue-shift by $\Delta\lambda \approx 100$ nm with the increase of laser fluence $F$, as shown in **Fig. 5**(b). However, such significant control of the optical resonances cannot be explained only by the measured small changes in geometry. In contrast to a previously demonstrated reshaping-based laser printing [40], we attribute the control to the intrinsic changes of the amorphous material. The difference in the refractive index of amorphous Si and crystalline Si is relatively large in the VIS spectral range, e.g. $\Delta n \approx 0.3$ at $\lambda = 532$ nm (see Appendix). The measured Raman spectra of the Si sample before and after the irradiation are plotted in **Fig. 5**(c). Raman measurements were performed using a commercially available confocal Raman system (WITec GmbH) equipped with a $\lambda = 785$ nm laser. The light was focused using a 50× objective (NA = 0.95) onto the sample and the Raman scattered light was collected with the same microscope objective. The laser power at the surface of the sample was $P = 1$ mW and the integration time of 1 s with 5 accumulations. Measured spectra were background corrected using the statistics-sensitive non-linear iterative peak-clipping (SNIP) algorithm with 100 iterations. For the template, a broad maximum is at $\tilde{\upsilon} = 480$ cm$^{-1}$, which corresponds to a vibrational mode of amorphous Si. A peak at $\tilde{\upsilon} = 510$ cm$^{-1}$ appears after the irradiation of the template and can be attributed to the formation of grain boundaries and nanocrystals, whereas the peak of crystalline Si is expected slightly further at $\tilde{\upsilon} = 520$ cm$^{-1}$ [57–59]. In addition, as the laser fluence is increased, the full-width at half maximum (FWHM) of the peak drops from $\Delta\tilde{\upsilon} \approx 53$ cm$^{-1}$ to $\Delta\tilde{\upsilon} \approx 20$ cm$^{-1}$. Therefore, the investigated phase transition to the polycrystalline phase is the main reason of the change in the optical response. See **Fig. 5**(d), for the obtained ED and MD resonances shift compared to an artificial case of a fixed refractive index and just the geometry change considered.

To this end, we have shown that we can modify the lithographically pre-processed nanostructure-based G filter to shift its optical resonances and subsequently change the color in transmission. Now, to emphasize the applicability more, we experimentally demonstrate, how the laser-induced thermal effects can be used to create user-defined patterns or separate pixels. First, using the same laser energies as in **Fig. 5**(b-d), we designed a unit of a multispectral filter array. In **Fig. 5**(e) we show experimentally realized transmitted colors, corresponding to the different laser energies applied, as depicted in a CIE 1931 chromaticity diagram. Furthermore, there we show a white-light transmission microscope image of a pixelated green filter with a pixel size of 50 μm. The high-precision of the laser beam permits a square-shape of the pixels and their accurate position in relation to each other. Each of the uniform pixels consists of ~450 partially overlapped laser shots. Despite the large number of laser shots (pulses), a laser source with a high repetition rate and a fast positioning table permits this within a matter of seconds. Moreover, even-though, the resolution of the laser pixelation in our proof-of-concept experiments was limited by the circular spot-size, it is sufficient to achieve square-like 5 μm pixels by a control of laser fluence in respect to the threshold and a partial overlap of 4 laser shots, as shown in the inset of **Fig. 5**(c). In future, the size of the laser spot could be reduced even further using an objective with a higher numerical aperture (NA) or a shorter irradiation wavelength $\lambda$, which would enable the use of the laser as a single nanostructure manipulation tool, as demonstrated by an atomic force microscope-assisted reshaping [35]. Besides the demonstrated pixelation of the G filter, the laser-induced tailoring via crystallization can be also applied to post-process nanostructure-based filters of other colors, e.g. red (R) and blue (B) filters – the standard RGB filters.

*3.2 Reshaping of Metasurface Elements Beyond Two-Dimensions*

In our experiments discussed so far, we excited the fundamental dipole resonances of the dielectric metasurface elements to tailor their refractive index without a firm control of the shape. The laser-induced reshaping of the nanostructures was initially demonstrated in plasmonics [34]. Later, a particular field distribution was used for an anisotropic reshaping of plasmonic nanostructures [35,38], but was never utilized in dielectric metasurfaces. Hence, we will demonstrate how by engineering the profile of the absorbed power and using ultrashort pulses one can obtain a spatially selective reshaping of dielectric metasurface elements.

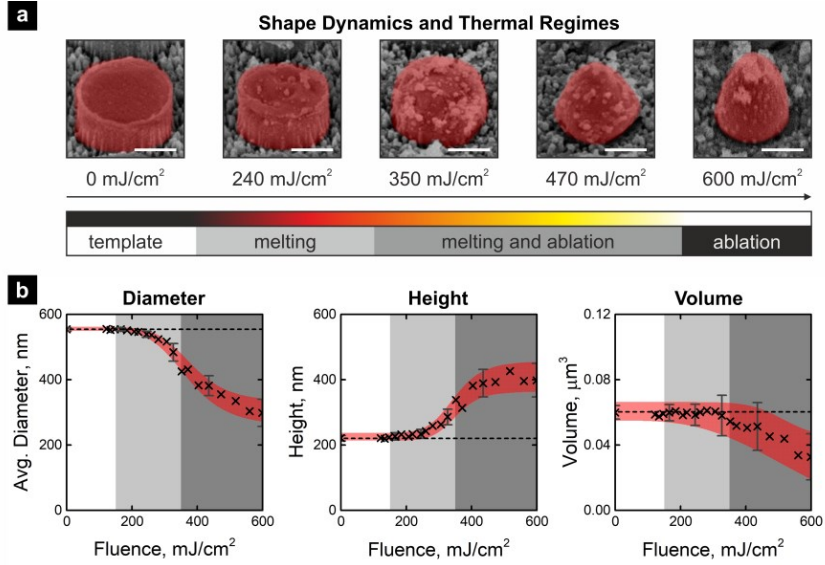

**Fig. 6.** Shape dynamics due to laser-induced tailoring of Si metasurface elements with ED and MD resonances in the IR. **(a)** SEM images of a typical unit cell after irradiation using different fluence and from the shape measurements identified thermal regimes. Insets show visual illustrations of the corresponding metasurface elements. **(b)** Measured geometrical parameters after irradiation: average diameter (left), height (center), and volume of the metasurface elements (right). Crosses depict measured values, and red shading denotes the interval of uncertainty. The dashed lines highlight the geometry values of the template. Melting as well as melting with ablation are highlighted in light- and dark grey, respectively.

For this matter, we selected a sample with the largest metasurface elements ($D = 560$ nm, $P = 794$ nm, $H = 220$ nm). Here, a laser irradiation at $\lambda = 532$ nm excites higher-order resonances and is absorbed predominantly at the top of metasurface elements, see **Fig. 2**(d,e). The template was irradiated by a set of single pulses while steadily increasing the energy until the point of ablation. The laser-induced thermal effects were analyzed by observation of the shape and the spectral response. In **Fig. 6**(a) we show SEM images of the metasurface elements irradiated by different laser energies, together with a scheme denoting the thermal regimes. For laser fluences below $F \approx 150$ mJ/cm² the geometry remains unchanged, while its further increase results in temperatures close to the melting point, $T_m = 1420$ K [56]. The nanostructures become round, but it is happening exclusively at the top of nanostructures. Such spatial confinement is attributed to the unique absorption profile with the hotspot at the top, see **Fig. 2**(e). The induced heat remains strongly localized during the 10 ps pulse duration, because it takes tens of nanoseconds for the heat to diffuse throughout the nanostructure. Subsequently, we note a substantial increase of the vertical dimensions of the modified nanostructures, while the diameter at the bottom seems to be sustained, as shown in **Fig. 6**(a). From the template up to the unstable case, the height $H$ increases almost by a factor of 2, from 220±13 nm to 434±54 nm, while the average diameter $D$ decreases from 560±5 nm to 300±42 nm, see **Fig. 6**(b). The nanostructure becomes more conical rather than hemispherical, as it was in the examples of continuous [39] or nanosecond pulse-induced heating [40].

Also, certain neglected aspects should be noted. First, the crystallization of Si is taking place, but the difference in the refractive index between amorphous and crystalline Si is negligible in the IR spectral range (see Appendix). Second, the heating leads to a decrease in the volume of the nanostructures starting at $F \approx 350$ mJ/cm², due to part of the sample reaching evaporation temperature, $T_v = 2654$ K [48]. Some of the evaporated material is redeposited on the surface, but the debris particles are significantly smaller than the elements of the metasurface, thus their influence on the optical response is also negligible in the investigated spectral range.

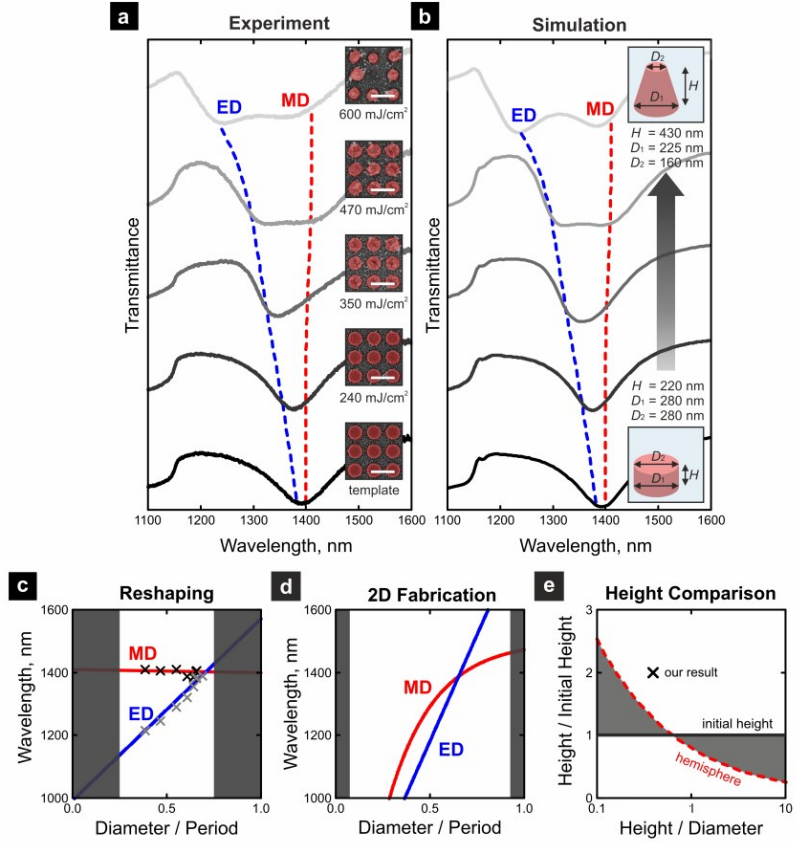

**Fig. 7**. Control of optical resonances based on metasurface reshaping beyond two-dimensions. (a) Transmittance measurements of an IR metasurface reshaped by increasing laser fluence, where ED and MD resonances are highlighted by the blue and red lines, respectively. The insets show SEM pictures of the measured arrays, the scale bars denote 1 μm. (b) Simulated transmittance of the arrays, with highlighted ED and MD resonances. Insets show the initial and the largest conical metasurface elements used. The structure is described by its height $H$, its diameter at the bottom $D_1$ and at the top $D_2$. (c) Measured spectral position of ED and MD resonances in dependence on the diameter scaled to the period for the sample analyzed in (a). The crosses denote measurement results; the lines are linear fits. (d) Simulated spectral position of ED and MD resonances in dependence on the diameter to period ratio for metasurfaces that could be realized by two-dimensional fabrication technologies for height $H$ = 220 nm and period $P$=794 nm. Black areas in (c,d) depict experimentally unattainable parameters. (e) Comparison of the obtained height to the generally predicted height transition from the initial to the hemispherical shape, plotted in respect to an aspect ratio.

Furthermore, the template and the reshaped samples were spectrally characterized using a broad-band spectrometer. As shown in **Fig. 7**(a), the dip in the transmission spectra, corresponding to the fundamental ED and MD resonances, tends to broaden and blue shift with the increase of the applied fluence $F$. As crystallization does not have a significant effect on the optical material properties of Si in the IR spectral range, this behavior can be directly related to the geometry dependent Mie-type electric and magnetic resonances. The subsequent optical simulations of the periodic conical metasurface elements provide identical spectral functions and confirm the positions of the fundamental ED and MD resonances, see **Fig. 7**(b). The diameter changes due to the laser irradiation. In **Fig. 7**(c), we plot the measured resonance positions versus the diameter $D$ scaled to the period $P$ of the metasurface. The ED resonance blue-shifts, while the MD resonance maintains its position. As the ED resonance is due to the collective polarization induced in the nanostructure, it is sensitive to the changes in the lateral dimensions, i.e. the diameter of the nanostructure. On other hand, the MD resonance is driven by the electric field coupling to displacement current loops, which are partially compensated by the increase of height.

Finally, to demonstrate the flexibility of the laser-induced reshaping for control of optical resonances, we also simulate the spectral positions of ED and MD resonances of dielectric metasurfaces with a fixed height of $H = 220$ nm, a period $P = 794$ nm, while varying a diameter $D$, within the limits allowed by fabrication using a standard two-dimensional lithography. The results are plotted in **Fig. 7**(d) in dependence on the diameter $D$ and they can be compared directly to the result of the laser-induced thermal reshaping in **Fig. 7**(c). In case of a fixed height, both ED and MD resonances blue-shift with the decrease of the diameter, whereas in case of laser-induced thermal reshaping the ED resonance can be varied independently of the MD, thus a prior unattainable control of optical resonances is obtained. Such control of the ED and MD resonances paves a way for a variety of applications, e.g. broadband band-stop filters and absorbers. Based on the spectral data shown in **Fig. 7**(a,b), the bandwidth of fundamental resonances can be flexibly adjusted and increased by as much as a factor of 2, from 120 nm to 240 nm. The largest experimentally obtained height is also significantly larger than predicted by the shape transition to a hemispherical shape, see **Fig. 7**(e).

On a side note, the demonstrated spatially-selective reshaping could also be applied to dielectric metasurfaces with their elements of different size, e.g. in case of the previously discussed smaller nanostructures with their resonances in VIS spectral range, by adjusting the wavelength of the laser irradiation.

## 4. Conclusions

The use of ultrashort pulse lasers enables a deterministic and spatially-selective tailoring of Si metasurfaces, thus has a great potential as a complementary technique for post-processing of large and uniform high-index dielectric metasurfaces.

First, we have demonstrated the applicability of the laser-induced thermal effects for a spatial variation of a large-scale metasurface, the so called laser pixelation, which is carried out by a precise positioning of the focused laser beam. The technique is mainly based on the crystallization of amorphous Si, which resulted in a $\Delta\lambda \approx 100$ nm blue-shift of the ED and MD resonances, sufficient to change the transmissive color of the nanostructure-based green filter to blue and, finally, red-shade. Such results indicate a simplification in the complex fabrication of the nanostructure-based color filter arrays, and a potential insight towards their implementation into daily-use digital imaging devices.

Second, we have introduced a multi-dimensional reshaping technique for dielectric nanostructures. We have shown, how a high absorption and a selection of a resonance-associated field distribution, allows to spatially engineer the hotspot for the reshaping, while a 10 ps pulse duration enables spatial confinement of the heating source. In the demonstrated example, we have shown an increase of nanostructure height by a factor of 2, while slightly shrinking its diameter. Such reshaping is followed by independent control of the ED and MD resonances, expanding the capabilities beyond the two-dimensional fabrication. The technique using ultrashort pulses is not surface energy limited as using continuous illumination or longer pulse durations. A great prospective is envisioned using even shorter pulses, spatial and temporal beam shaping approaches.

In addition, the laser-induced thermal effects were shown dependent on Si parameters, but are transferrable to metasurfaces of other high-index materials, e.g. $TiO_2$, Ge, etc., while the selection of an ultrashort pulse laser with a high repetition rate as a light source gives the opportunity not only for the spatial but also for a rapid and efficient post-processing, which suggests a direct path towards a mass-production of rigid metasurface-based optical elements.

**Appendix: Refractive Index**

The VIS samples were made from a commercial amorphous Si layer (Tafelmaier Dünnschicht-Technik GmbH), see **Fig. 8**(a) for measured refractive index in comparison to crystalline Si data from literature [60]. The IR sample was structured from amorphous Si film obtained via electron beam evaporation. Its real part of refractive index was taken from commercial amorphous Si, while extinction was fitted via comparison of measured and simulated transmittance, see dispersion parameters in **Fig. 8**(b).

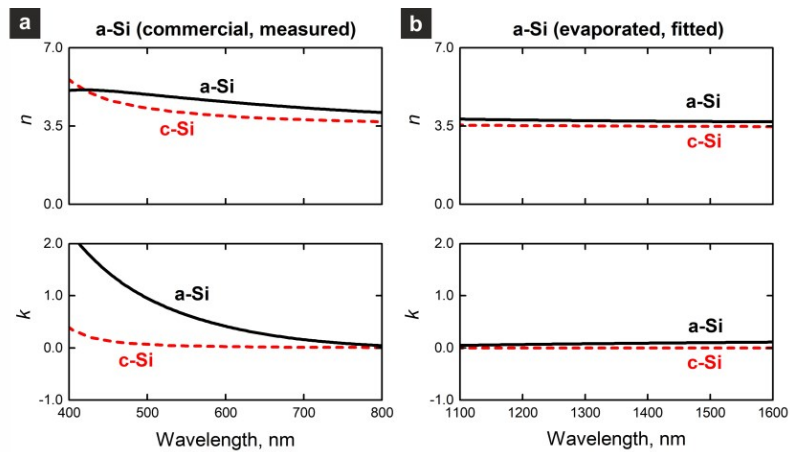

**Fig. 8.** Refractive index $n$, $k$ of different Si samples: (a) measured data of commercial amorphous Si used for VIS samples, (b) fitted data of amorphous Si deposited by electron beam evaporation, used for IR sample. Red dashed line depicts refractive index of crystalline Si from [60].

**Funding**

European Union's Horizon 2020 research and innovation programme under the Marie Sklodowska-Curie grant agreement No. 675745; German Federal Ministry of Education and Research (FKZ 03ZZ0434, FKZ 03Z1H534, FKZ 03ZZ0451).

**Acknowledgments**

The authors thank Isabelle Staude from Friedrich Schiller University Jena for helpful discussions and IR sample, Dennis Arslan from Friedrich Schiller University Jena for technical assistance in spectral measurements, and Arvind Nagarajan from TNO for useful suggestions.